\PassOptionsToPackage{bookmarks={false}}{hyperref}
\documentclass[sigconf]{acmart}

\acmConference[]{}{}
\acmBooktitle{}
\renewcommand\footnotetextcopyrightpermission[1]{}
\usepackage{natbib}
\usepackage{multibib}
\newcites{web}{Online Artefacts}
\usepackage{cleveref}
\def\NumPostsFromSO{1,425}

\usepackage{xcolor}

\usepackage{tablefootnote}
\begin{document}
\pagestyle{plain}
\title{Ranking Computer Vision Service Issues using Emotion}

\author{Maheswaree K Curumsing}
\email{m.curumsing@deakin.edu.au}
\affiliation{ \department{Applied Artificial Intel. Inst.}
 \institution{Deakin University}
 \streetaddress{Burwood}
 \city{Geelong}
 \state{Victoria}
 \country{Australia}
 }

\author{Alex Cummaudo}
\email{ca@deakin.edu.au}
\affiliation{ \department{Applied Artificial Intel. Inst.}
 \institution{Deakin University}
 \streetaddress{Burwood}
 \city{Geelong}
 \state{Victoria}
 \country{Australia}}
 \author{Ulrike Maria Graetsch}
\email{maria.graetsch@deakin.edu.au}
\affiliation{ \department{Applied Artificial Intel. Inst.}
 \institution{Deakin University}
 \streetaddress{Burwood}
 \city{Geelong}
 \state{Victoria}
 \country{Australia}}
 \author{Scott Barnett}
\email{scott.barnett@deakin.edu.au}
\affiliation{ \department{Applied Artificial Intel. Inst.}
 \institution{Deakin University}
 \streetaddress{Burwood}
 \city{Geelong}
 \state{Victoria}
 \country{Australia}}
 \author{Rajesh Vasa}
\email{rajesh.vasa@deakin.edu.au}
\affiliation{ \department{Applied Artificial Intel. Inst.}
 \institution{Deakin University}
 \streetaddress{Burwood}
 \city{Geelong}
 \state{Victoria}
 \country{Australia}}
 \renewcommand{\shortauthors}{Curumsing et al.}

\begin{abstract}
Software developers are increasingly using machine learning APIs to implement `intelligent' features. Studies show that incorporating machine learning into an application increases technical debt, creates data dependencies, and introduces uncertainty due to non-deterministic behaviour. However, we know very little about the emotional state of software developers who deal with such issues. In this paper, we do a landscape analysis of emotion found in 1,245 Stack Overflow posts about computer vision APIs. We investigate the application of an existing emotion classifier EmoTxt and manually verify our results. We found that the emotion profile varies for different question categories.

\end{abstract}

\maketitle
\section{Introduction}
Recent advances in artificial intelligence have provided software engineers with new opportunities to incorporate complex machine learning capabilities, such as computer vision, through cloud based `intelligent' web services. These new set of services, typically offered as API calls are marketed as a way to reduce the complexity involved in integrating AI-components. However, recent work shows that software engineers struggle to use these intelligent services~\citep{Cummaudo:2019vi}.
Furthermore, the accompanying documentation fails to address common issues experienced by software engineers and, often, engineers resort to online communication channels, such as, JIRA and Stack Overflow (SO) to seek advice from their peers~\citep{Cummaudo:2019vi}.
While seeking advice on the issues, software engineers tend to express their emotions (such as frustration or confusion) within the questions. Recognising the value of considering emotions, other researchers have investigated emotions expressed by software developers within communication channels~\citep{ortu2016} including Stack Overflow (SO)~\citep{novielli2018, calefato2018}; the broad motivation of these works is to generally understand the emotional landscape and improve developer productivity~\citep{murgia2014, ortu2016, gachechiladze2017}. However, previous works have not directly focused on the nature of emotions expressed in questions related to intelligent web services. We also do not know if certain types of questions express stronger emotions.
The machine-learnt behaviour of these cloud intelligent services is typically non-deterministic and, given the dimensions of data used, their internal inference process is hard to reason about~\citep{Cummaudo:2019va}. Compounding the issue, documentation of these cloud systems does not explain the limits, nor how they were created (esp. data sets used to train them). This lack of transparency makes it difficult for even senior developers to properly reason about these systems, so their prior experience and anchors do not offer sufficient support~\citep{Cummaudo:2019vi}. In addition, adding machine learned behaviour to a system incurs ongoing maintenance concerns~\citep{sculley2015hidden}. There is a need to better understand emotions expressed by developers to inform cloud vendors and help them improve their documentation and error messages provided by their services.
This work builds on top of recent work that explored \textit{what} pain-points developers face when using intelligent services through a general analysis of \NumPostsFromSO{} SO posts (questions)~\citep{Cummaudo:2019vi} using an existing SO issue classification taxonomy~\citep{Beyer:2018fm}. In this work, we consider the emotional state expressed within these pain-points, using the same data set of \NumPostsFromSO{} SO posts. We identify the emotions in each SO question, and investigate if the distribution of these emotions is similar across the various types of questions.
In order to classify emotions from SO posts, we use EmoTxt, a recently proposed toolkit for emotion recognition from text~\citep{novielli2018, calefato2017, calefato2018}. EmoTxt has been trained and built on SO posts using the emotion classification model proposed by~\citet{shaver1987}. The category of issue was manually determined in our prior work.

The key findings of our study are:
\begin{itemize}
    \item The distribution of emotions is different across the taxonomy of issues.
    \item A deeper analysis of the results, obtained from the EmoTxt classifier, suggests that the classification model needs further refinement. Love and joy, the least expected emotions when discussing API issues, are visible across all categories.
    \end{itemize}
    In order to promote future research and permit replication, we make our data set publicly available.\footnote{See \url{http://bit.ly/2RiULgW}.} The paper structure is as follows: \cref{sec:EM} provides an overview on prior work surrounding the classification of emotions from text; \cref{sec:methodology} describes our research methodology;  \cref{sec:findings} presents the results from the EmoTxt classifier;  \cref{sec:discussion} provides a discussion of the results obtained;
\cref{sec:threats} outlines the threats to validity; \cref{sec:conclusion} presents the concluding remarks.

\section{Emotion Mining from Text}\label{sec:EM}
Several studies have investigated the role of emotions generally in software development~\citep{wrobel2013, Shaw:2003aa, ortu2016, gachechiladze2017}. Work in the area of behavioural software engineering established the link between software developer's happiness and productivity~\citep{Graziontin:2017}. Wrobel~\citep{wrobel2013} investigated the impact that software developers' emotion has on the development process and found that frustration and anger were amongst the emotions that posed the highest risk to developer's productivity.
Recent studies focused on emotion mining from text within communication channels used by software engineers to communicate with their peers~\citep{murgia2014, ortu2016, gachechiladze2017, novielli2018}.~\citet{murgia2014} and~\citet{ortu2016} investigated the emotions expressed by developers within an issue tracking system, such as JIRA, by labelling issue comments and sentences written by developers using Parrott's framework.~\citet{gachechiladze2017} applied the Shaver framework to detect anger expressed in comments written by developers in JIRA. The Collab team~\citep{calefato2017, novielli2018} extended the work done by~\citet{ortu2016} and developed a gold standard data set collected from SO posts consisting of questions, comments and feedback. This data set was manually annotated using the Shaver's emotion model. The Shaver's model consists of a tree-structured, three level, hierarchical classification of emotions. The top level consists of six basic emotions namely, love, joy, anger, sadness, fear and surprise~\citep{shaver1987}. The subsequent levels further refines the granularity of the previous level. One of their recent work~\citep{novielli2018} involved 12 raters to manually annotate 4,800 posts (where each post included the question, answer and comments) from SO. The same question was assigned to three raters to reduce bias and subjectivity. Each coder was requested to indicate the presence/absence of each of the six basic emotions from the Shaver framework. As part of their work they developed an emotion mining toolkit, EmoTxt~\citep{calefato2017}. The work conducted by the Collab team is most relevant to our study since their focus is on identifying emotion from SO posts and their toolkit is trained on a large data set of SO posts.
\section{Methodology}\label{sec:methodology}
As mentioned in our introduction, this paper uses the data set reported in~\citeauthor{Cummaudo:2019vi}'s ICSE 2020 paper~\citep{Cummaudo:2019vi}. As this paper is in press, we reproduce a summary of the methodology used in constructing this data set methodology below. For full details, we refer to the original paper. Supplementary materials used for this work are provided for replication.\footnotemark[1]
Our research methodology consisted of the following steps: (i)~data extraction from Stack Overflow resulting in 1,425 questions about intelligent computer vision services; (ii)~question classification using the taxonomy presented by \citet{Beyer:2018fm}; (iii)~automatic emotion classification using EmoTxt based on \citeauthor{shaver1987}'s emotion taxonomy \citep{shaver1987}; and (iv)~manual classification of 25 posts to better understand developers emotion. We calculated the inter-rater reliability between EmoTxt and our manually classified questions in two ways: (i)~to see the overall agreement between the three raters in applying the \citeauthor{shaver1987} emotions taxonomy, and (ii)~to see the overall agreement with EmoTxt's classifications. Further details are provided below.
\def \alexnumauthor {second}
\subsection{Data Set Extraction from SO}
\subsubsection{Intelligent Service Selection}
We contextualise this work within popular computer vision service providers: Google Cloud~\citepweb{GoogleCloud:Home}, AWS~\citepweb{AWS:Home}, Azure~\citepweb{Azure:Home} and IBM Cloud~\citepweb{IBM:Home}.
We chose these four providers given their prominence and ubiquity as cloud service vendors, especially in enterprise applications~\citep{RightScaleInc:2018kJ}. We acknowledge other services beyond the four analysed which provide similar capabilities~\citepweb{Pixlab:Home,Clarifai:Home,Cloudsight:Home,DeepAI:Home,Imagaa:Home,Talkwaler:Home}. Additionally, only English-speaking services have been selected, excluding popular computer vision services from Asia (e.g.,~\citepweb{Megvii:Home,TupuTech:Home,YiTuTech:Home,SenseTime:Home,DeepGlint:Home}).
\subsubsection{Developing a search query}
To understand the various ways developers refer to these services, we needed to find search terms that are commonplace in question titles and bodies that discuss the service names. One approach is to use the \textit{Tags} feature in SO. To discover which tags may be relevant, we ran a search\footnote{The query was run on January 2019.} within SO against the various brand names of these computer vision services, reviewed the first three result pages, and recorded each tag assigned per question.\footnote{Up to five tags can be assigned per question.} However, searching using tags alone on SO is ineffective (see~\citep{Tahir:2018ks,Barua:2012gz}).
To overcome this limitation, we ran a second query within the Stack Exchange Data Explorer\footnote{\url{http://data.stackexchange.com/stackoverflow}} (SEDE) using these tags, we sampled 100 questions (per service), and noted the permutations in how developers refer to each service\footnote{E.g., misspellings, misunderstanding of brand names, hyphenation, UK vs. US English, and varied uses of apostrophes, plurals, and abbreviations.}. We noted 229 permutations.
\subsubsection{Executing our search query}
Next, we needed to extract questions that make reference to any of these 229 permutations. SEDE has a 50,000 row limit and does not support case-insensitivity, however Google's BigQuery does not. Therefore, we queried Google's SO dataset on each of the 229 terms that may occur within the title or body of question posts,\footnote{See \url{http://bit.ly/2LrN7OA}.} which resulted in 21,226 questions.
\subsubsection{Refining our inclusion/exclusion criteria}
\label{ssec:method:filtering:refining}
To assess the suitability of these questions, we filtered the 50 most recent posts as sorted by their \textit{CreationDate} values. This helped further refine the inclusion and exclusion criteria: for example, certain abbreviations in our search terms (e.g., `GCV', `WCS'\footnote{Watson Cognitive Services}) allowed for false positive questions to be included, which were removed. Furthermore, we consolidated all overlapping terms (e.g., `Google Vision \underline{\textbf{API}}' was collapsed into `Google Vision') to enhance the query. Additionally, we reduced our 221 search terms to just 27 search terms by focusing on computer vision services \textit{only}\footnote{Our original data set aimed at extracting posts relevant to \textit{all} intelligent services, and not just computer vision services. However, 21,226 questions were too many to assess without automated analysis, which was beyond the scope of our work.} which resulted in \NumPostsFromSO{} questions. No duplicates were recorded as determined by the unique ID, title and timestamp of each question.
\subsubsection{Manual filtering}
\label{ssec:method:filtering:automated-manual-filtering}
The next step was to assess the suitability and nature of the \NumPostsFromSO{} questions extracted. The \alexnumauthor{} author ran a manual check on a random sample of 50 posts, which were parsed through a templating engine script\footnote{We make this available for future use at: \url{http://bit.ly/2NqBB70}.} in which the ID, title, body, tags, created date, and view, answer and comment counts were rendered for each post.
Any match against the 27 search terms in the title or body of the post were highlighted, in which three false positives were identified as either library imports or stack traces, such as \texttt{aws-java-sdk-\underline{rekognition}:jar}. In addition, we noted that there were false positive hits related to non-computer vision services. We flagged posts of such nature as `noise' and removed them from further classification.
\begin{table*}[tb]
  \centering
  \caption{Descriptions of dimensions from our interpretation of \citeauthor{Beyer:2018fm}'s SO question type taxonomy.}
  \label{tab:taxonomies}
  \small
  \begin{tabular}{p{.11\linewidth}p{.83\linewidth}}
    \toprule
        \textbf{Dimension} & \textbf{Our Interpretation}
    \\
    \midrule

    \textbf{API usage\dotfill} &
    Issue on how to implement something using a specific component provided by the API
    \\
        \textbf{Discrepancy\dotfill} &
    The questioner's \textit{expected behaviour} of the API does not reflect the API's \textit{actual behaviour}
    \\
        \textbf{Errors\dotfill} &
    Issue regarding an error when using the API, and provides an exception and/or stack trace to help understand why it is occurring
    \\
        \textbf{Review\dotfill} &
    The questioner is seeking insight from the developer community on what the best practices are using a specific API or decisions they should make given their specific situation
    \\
        \textbf{Conceptual\dotfill} &
    The questioner is trying to ascertain limitations of the API and its behaviour and rectify issues in their conceptual understanding on the background of the API's functionality
    \\
        \textbf{API change\dotfill} &
    Issue regarding changes in the API from a previous version
    \\
        \textbf{Learning\dotfill} &
    The questioner is seeking for learning resources to self-learn further functionality in the API, and unlike discrepancy, there is no specific problem they are seeking a solution for
    \\
    \bottomrule
  \end{tabular}
  \vspace{-2mm}
\end{table*}
\subsection{Question Type \& Emotion Classification}
\subsubsection{Manual classification of question category}
\label{ssec:method:filtering:classification}
We classify our \NumPostsFromSO{} posts using ~\citeauthor{Beyer:2018fm}'s taxonomy~\citep{Beyer:2018fm} as it was comprehensive and validated \citep{Cummaudo:2019vi}. We split the posts into 4 additional random samples, in addition to the random sample of 50 above. 475 posts were classified by the \alexnumauthor{} author and three other research assistants\footnote{Software engineers in our research group with at least 2 years industry experience} classified the remaining 900 (i.e., a total of 1,375 classifications). An additional 450 classifications were assigned due to reliability analysis, in which the remaining 50 posts were classified nine times by various researchers in our group.\footnote{Due to space limitations, reliability analysis is omitted and is reported in~\citep{Cummaudo:2019vi}.}
Due to the nature of reliability analysis, multiple classifications (450) existed for  these 50 posts. Therefore, we applied a `majority rule' technique to each post allowing for a single classification assignment and therefore analysis within our results. When there was a majority then we used the majority classification; when there was a tie, then we used the classification that was assigned the most out of the entire 450 classifications. As an example, 3 raters classified a post as \textit{API Usage}, 1 rater classified the same post as a \textit{Review} question and 5 raters classified the post as \textit{Conceptual}, resulting in the post being classified as a \textit{Conceptual} question. For another post, three raters assigned \textit{API Usage}, \textit{Discrepancy} and \textit{Learning} (respectively), while 3 raters assigned \textit{Review} and 3 raters assigned \textit{Conceptual}. In this case, \textit{Review} and \textit{Conceptual} were tied, but was resolved down to \textit{Conceptual} as this classification received 147 more votes than \textit{Review} across all classifications made in the sample of 50 posts.
However, where a post was extracted from our original \NumPostsFromSO{} posts but was either a false positive, not applicable to intelligent services (see \cref{ssec:method:filtering:automated-manual-filtering}), or not applicable to a taxonomy dimension/category, then the post was flagged for removal in further analysis. This was done 180 times, leaving a total of 1,245 posts.

\subsubsection{Emotion classification using ML techniques}
After extracting and classifying all posts, we then piped in the body of each question into a script developed to remove all HTML tags, code snippets, blockquotes and hyperlinks, as suggested by~\citet{novielli2018}. We replicated and extended the study conducted by~\citet{novielli2018} on our data set derived from \NumPostsFromSO{} SO posts, consisting of questions only. Our study consisted of three main steps, namely, (1) automatic emotion classification using EmoTxt, (2) manual annotation process and, (3) comparison of the automatic classification result with the manually annotated data set.
\subsubsection{Emotion classification using EmoTxt}
We started with a file containing 1,245 non-noise SO questions, each with an associated question type as classified using the strategy discussed in \cref{ssec:method:filtering:classification}. We pre-processed this file by extracting the question ID and body text to meet the format requirements of the EmoTxt classifier~\citep{calefato2017}. This classifier was used as it was trained on SO posts as discussed in Section \ref{sec:EM}. We ran the classifier for each emotion as this was required by EmoTxt model.  This resulted in 6 output prediction files (one file for each emotion: \textit{Love}, \textit{Joy}, \textit{Surprise}, \textit{Sadness}, \textit{Fear}, \textit{Anger}). Each question within these files referenced the question ID and a predicted classification (\texttt{YES} or \texttt{NO}) of the emotion.  We then merged the emotion prediction files into an aggregate file with question text and~\citeauthor{Beyer:2018fm}'s taxonomy classifications. This resulted in 796 emotion classifications. We further analysed the classifications and generated an additional classification of \textit{No Emotion} for the 622 questions where EmoTxt predicted \texttt{NO} for all the emotion classification runs.
Of the 796 questions with emotion detected, 143 questions had 2 or more emotions predicted: 1 question\footnote{See \url{http://stackoverflow.com/q/55464541}.} had up to 4 emotions detected (\textit{Surprise}, \textit{Sadness}, \textit{Joy} and \textit{Fear}), 28 questions had up to 3 emotions detected, and the remaining 114 had up to two emotions detected.
\def\light{$L_{\kappa}$}
\subsubsection{Manual Annotation Process} \label{ssec:manual}
In order to evaluate and also better understand the process used by EmoTxt to classify emotions, we manually annotated a small sample of 25 SO posts, randomly selected from our data set. Each of these 25 posts were assigned to three raters who carried out the following three steps: (i) identify the presence of an emotion; (ii) if an emotion(s) exists, classify the emotion(s) under one of the six basic emotions proposed by the Shaver framework~\citep{shaver1987}; (iii) if no emotion is identified, annotate as neutral.
We then collated all rater's results and calculated Light's Kappa (\light{})~\citep{Light:1971vz} to measure the overall agreement \textit{between} raters to measure the similarity in which independent raters classify emotions to SO posts. As \light{} does not support multi-class classification (i.e., multiple emotions) per subjects (i.e., per SO post), we binarised the results each emotion and rater as \texttt{TRUE} or \texttt{FALSE} to indicate presence, calculated the \light{} per emotion against the three raters, and averaged the result across all emotions to get an overall strength of agreement.
\subsubsection{Comparing EmoTxt results with the results from Manual Classification}
\def\cohen{$C_{\kappa}$}
The next step involved comparing the ratings of the 25 SO posts that were manually annotated by the three raters with the results obtained for the same set of 25 SO posts from the EmoTxt classifier.
Similar to \cref{ssec:manual}, we used Cohen's Kappa (\cohen{})~\citep{Cohen:1960tf} to measure the consistency of classifications of EmoTxt's classifications versus the manual classifications of each rater. We separated the classifications per emotion and calculated \cohen{} for each rater against EmoTxt and averaged these values for all emotions.
After noticing poor results, the three raters involved in \cref{ssec:manual} were asked to compare and discuss the ratings from the EmoTxt classifier against the manual ratings.
The findings from this process are presented and discussed in the next two sections.

\section{Findings}\label{sec:findings}
\begin{figure}[h]
\includegraphics[width=1\linewidth]{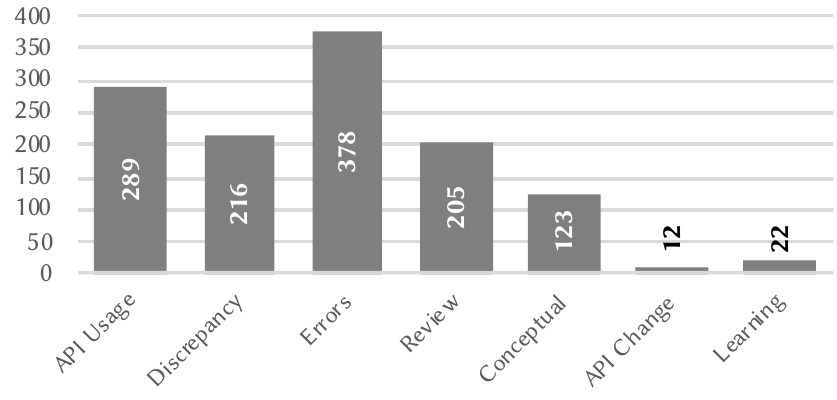}
\caption{Distribution of SO question types}
\label{fig:beyer-classifications}
\end{figure}
\begin{figure}[tbh]
\includegraphics[width=\linewidth]{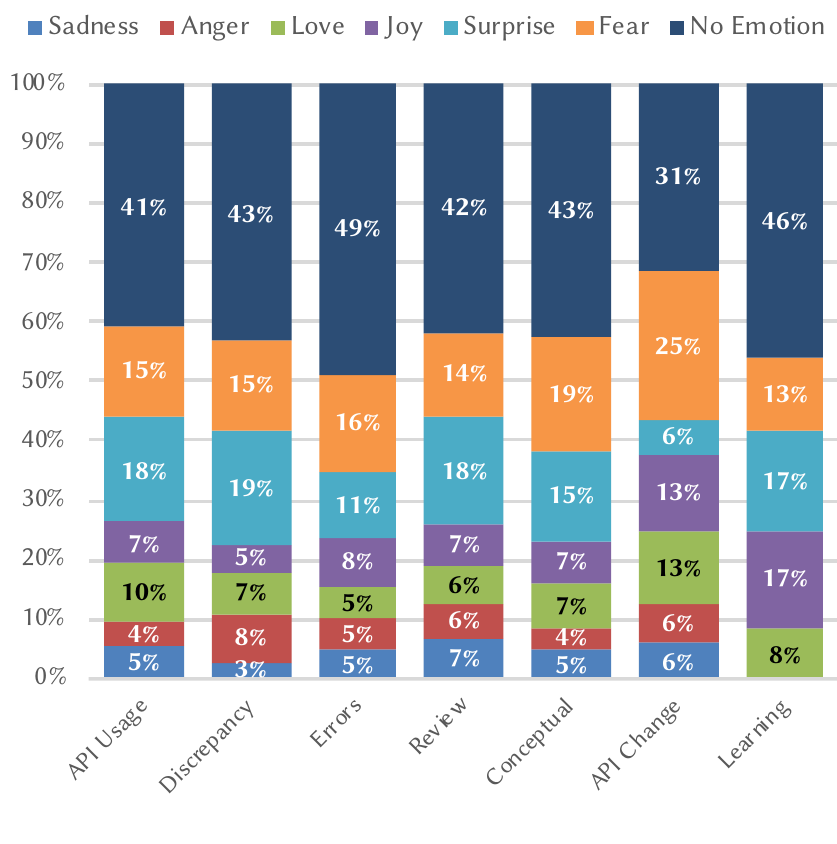}
\vspace{-3.5\bigskipamount}
\caption{Proportion of emotions per question type.}
\vspace{-2\bigskipamount}
\label{fig:emotion-dist}
\end{figure}

\Cref{fig:beyer-classifications} displays the overall distribution of question types from the 1,245 posts classified in~\citep{Cummaudo:2019vi}, when adjusted for majority ruling as per \cref{ssec:method:filtering:classification}. It is evident that developers ask issues predominantly related to API errors when using computer vision services and, additionally, how they can use the API to implement specific functionality. There are few questions related to version issues or self-learning.
\begin{table}[th]
\caption{Frequency of emotions per question type.}
\label{tab:emotion-freq}
\resizebox{\linewidth}{!}{\begin{tabular}{l|ccccccc|c}
\toprule
\textbf{Question Type}&
\textbf{Fear}&
\textbf{Joy}&
\textbf{Love}&
\textbf{Sadness}&
\textbf{Surprise}&
\textbf{Anger}&
\textbf{No Emotion}&
\textbf{Total}\\
\midrule
API Usage&50&22&34&18&59&13&135&331\\
Discrepancy&38&12&18&7&48&20&108&251\\
Errors&69&34&22&21&48&23&206&423\\
Review&34&16&15&16&42&14&98&235\\
Conceptual&26&10&10&7&21&5&59&138\\
API Change&4&2&2&1&1&1&5&16\\
Learning&3&4&2&0&4&0&11&24\\
\midrule
Total&224&100&103&70&223&76&622&1418\\
\bottomrule
\end{tabular}}
\end{table}
\Cref{tab:emotion-freq} displays the frequency of questions that were classified by EmoTxt when compared to our assignment of question types, while \cref{fig:emotion-dist} presents the emotion data proportionally across each type of question. \textit{No Emotion} was the most prevalent across all question types, which is consistent with the findings of the Collab group during the training of the EmoTxt classifier. Interestingly, \textit{API Change} questions had a distinct distribution of emotions, where 31.25\% of questions had \textit{No Emotion} compared to the average of 42.01\%. This is likely due to the low sample size of \textit{API Change} questions, with only 12 assignments, however the next highest set of emotive questions are found in the second largest sample (\textit{API Usage}, at 59.21\%) and so greater emotion detected is not necessarily proportional to sample size.  Unsurprisingly, \textit{Discrepancy} questions had the highest proportion of the \textit{Anger} emotion, at 7.97\%, compared to the mean of 4.74\%, which is indicative of the frustrations developers face when the API does something unexpected. \textit{Love}, an emotion which we expected least by software developers when encountering issues, was present across the different question types. The two highest emotions, by average, were \textit{Fear} (16.67\%) and \textit{Surprise} (14.90\%), while the two lowest emotions were \textit{Sadness} (4.47\%) and \textit{Anger} (4.74\%). \textit{Joy} and \textit{Love} were roughly the same and fell in between the two proportion ends, with means of 8.96\% and 8.16\%, respectively.
Results from our reliability analysis showed largely poor results. Guidelines of indicative strengths of agreement are provided by~\citet{Landis:1977kv}, where $\kappa \leq 0.000$ is \textit{poor agreement}, $0.000 < \kappa \leq 0.200$ is \textit{slight agreement} and $0.200 < \kappa \leq 0.400$ is \textit{fair agreement}. Our readings were indicative of poor agreement between raters (\cohen{}$=-0.003$) and slight agreement with EmoTxt (\light{}$=0.155$).  The strongest agreements found were for \textit{No Emotion} both between each of our three raters (\light{}$=0.292$) and each rater and EmoTxt (\cohen{}$=0.086$), with fair and slight agreement respectively.
\begin{table*}
\caption{Sample questions comparing question type to emotion. Questions located at https://stackoverflow.com/q/[ID].}
\label{table:SampleQuestions}
\small
\begin{tabular}{lp{0.71\linewidth}ll}
\toprule
\textbf{ID} &
\textbf{Quote}&
\textbf{Classification}&
\textbf{Emotion}\\
\midrule
53249139&
\textit{``I'm trying to integrate my project with Google Vision API... I'm wondering if there is a way to set the credentials explicitly in code as that is more convenient than setting environment variables in each and every environment we are running our project on... I know for a former client version 1.22 that was possible... but for the new client API I was not able to find the way and documentation doesn't say anything in that regards.''}&
API Usage&
Fear\\
40013910&
\textit{``I want to say something more about Google Vision API Text Detection, maybe any Google Expert here and can read this. As Google announced, their TEXT\_DETECTION was fantastic... But for some of my pics, what happened was really funny... There must be something wrong with the text detection algorithm.''}&
Discrepancy&
Anger\\
50500341&
\textit{``I just started using PYTHON and now i want to run a google vision cloud app on the server but I'm not sure how to start. Any help would be greatly appreciated.''}&
API Usage&
Sadness\\
49466041&
\textit{``I am getting the following error when trying to access my s3 bucket... my hunch is it has something to do with the region...I have given almost all the permissions to the user I can think of.... Also the region for the s3 bucket appears to be in a place that can work with rekognition. What can I do?''}&
Errors&
Surprise\\
55113529&
\textit{''Following a tutorial, doing everything exactly as in the video... Hoping to figure this out as it is a very interesting concept...Thanks for the help... I'm getting this error:...''}&
Errors&
Joy\\
39797164&
\textit{``Seems that the Google Vision API has moved on and the open Sourced version has not....In my experiments this `finds' barcodes much faster than using the processor that the examples show.  Am I missing something somewhere?''}&
API Change&
Love\\
\bottomrule
\end{tabular}
\end{table*}

\section{Discussion}\label{sec:discussion}
Our findings from the comparison between the manually annotated SO posts and the automatic classification revealed substantial discrepancies. \Cref{table:SampleQuestions} provide some sample questions from our data set and the emotion identified by EmoTxt within the text. A subset of questions analysed by our three raters do not indicate the automatic (EmoTxt) emotion, and upon manual inspection of the text after poor results from our reliability analysis, an introspection of the data set sheds some light to the discrepancy. For example, question 55113529 shows no indication of \textit{Joy}, rather the developer is expressing a state of confusion. The phrase \textit{``Thanks for your help''} could be the reason why the miss-classification occurred if words like ``thanks'' were associated with joy. However, in this case, it seems unlikely that the developer is expressing joy as the developer has followed a tutorial but is still encountering an error. Similarly, question 39797164, classified as \textit{Love} and question 50500341, classified as \textit{Sadness} express a state of confusion and the urge to know more about the product; upon inspecting the entire question in context, it is difficult to consistently agree with the emotions as determined by EmoTxt, and further exploration into the behaviour and limitations of the model is necessary.
Our results indicate further work is needed to refine the ML classifiers that mine emotions in the SO context. The question that arises is whether the classification model is truly reflective of real-world emotions expressed by software developers. As highlighted by~\citet{curumsing2017}, the divergence of opinions with regards to the emotion classification model proposed by theorists raises doubts to the foundations of basic emotions. Most of the studies conducted in the area of emotion mining from text is based on an existing general purpose emotion framework from psychology~\citep{Ondrej:2016, ortu2016, novielli2018} -- none of which are tuned for software engineering domain. In our our study, we note the emotions expressed by software developers within SO posts are quite narrow and specific. In particular, emotions such as frustration and confusion would be more appropriate over love and joy.

\section{Threats to Validity}\label{sec:threats}
\paragraph{Internal validity:}
The \textit{API Change} and \textit{Learning} question types were few in sample size (only 12 and 22 questions, respectively). The emotion  proportion distribution of these question types are  quite different to the others.  Given the low number of questions, the sample is too small to make confident assessments. Furthermore, our assignment of~\citeauthor{Beyer:2018fm}'s question type taxonomy was single-label; a multi-labelled approach may work better, however analysis of results would become more complex. A multi-labelled approach would be indicative for future work.
\paragraph{External validity:}
EmoTxt was trained on questions, answers and comments, however our data set contained questions only. It is likely that our results may differ if we included other discussion items, however we wished to understand the emotion within developers' \textit{questions} and classify the question based on the question classification framework by~\citet{Beyer:2018fm}. Moreover, this study has only assessed frustrations within the context of a concrete domain; intelligent computer vision services. The generalisability of this study to other intelligent services, such as natural language processing services, or conventional web services, may be different. Furthermore, we only assessed four popular computer vision services; expanding the data set to include more services, including non-English ones, would be insightful. We leave this to future work.
\paragraph{Construct validity:}
Some posts extracted from SO were false positives. Whilst flagged for removal (\cref{ssec:method:filtering:automated-manual-filtering}), we cannot guarantee that all false positives were removed. Furthermore, SO is known to have questions that are either poorly worded or poorly detailed, and developers sometimes ask questions without doing any preliminary investigation. This often results in down-voted questions. We did not remove such questions from our data set, which may influence the measurement of our results.

\section{Conclusion}\label{sec:conclusion}
In this paper we analysed SO posts for emotions using an automated tool and cross-checked it manually. We found that the distribution of emotion differs across the taxonomy of issues, and that the current emotion model typically used in recent works is not appropriate for emotions expressed within SO questions. Consistent with prior work~\citep{lin2018sentiment}, our results demonstrate that machine learning classifiers for emotion are insufficient; human assessment is required.

\begin{anonsuppress}
\section*{Acknowledgements}
We acknowledge the efforts of Darren Annal who assisted with classifying Stack Overflow posts with the emotion taxonomy.
\end{anonsuppress}
\bibliography{references}
\bibliographystyleweb{IEEEtran}
\bibliographyweb{web}
\end{document}